\documentclass{PoS}
\usepackage{tabularx}
\usepackage{graphicx}
\usepackage{epsfig} 

\title{The Sphaleron Rate at the Electroweak Crossover}

\ShortTitle{The sphaleron rate at the electroweak crossover}

\author{\speaker{Michela D'Onofrio}\\
        Department of Physics, University of Helsinki and Helsinki Institute of Physics\\
        P.O. Box 64, FI-00014 Helsinki\\
        E-mail: \email{michela.donofrio@helsinki.fi}}

\author{Kari Rummukainen\\
		Department of Physics, University of Helsinki and Helsinki Institute of Physics\\
        P.O. Box 64, FI-00014 Helsinki\\        
        E-mail: \email{kari.rummukainen@helsinki.fi}}

\author{Anders Tranberg\\
        Niels Bohr Institute, University of Copenhagen\\
        Blegdamsvej 17, DK-2100 Copenhagen\\
        and Helsinki Institute of Physics\\
        P.O. Box 64, FI-00014 Helsinki\\ 
        E-mail: \email{antranbe@nbi.dk}}

\abstract{The baryon number is violated in the Standard Model by non-perturbative sphaleron transitions. At temperatures above the electroweak scale, the rate of the sphaleron transitions is unsuppressed and has been accurately measured using effective theories on the lattice. At temperatures substantially below the electroweak scale, the Higgs field expectation value is large and the sphaleron rate is strongly suppressed. Here analytical estimates are sufficient. The sphaleron rate, however, has not been calculated in the intermediate temperature range with physical Standard Model parameters. In this work we use an effective electroweak theory on the lattice with multicanonical and real-time simulation methods to calculate the sphaleron rate through the electroweak crossover at Higgs masses of 115 GeV and 160 GeV. The results are significant e.~g.~for Leptogenesis scenarios.

\vspace{-166mm}\parbox{\textwidth}{\flushright\large\rm \hfill HIP-2010-30/TH}\vspace{160mm}
}

\FullConference{The XXVIII International Symposium on Lattice Field Theory, Lattice2010\\
		June 14-19, 2010\\
		Villasimius, Italy}

\begin{document}

\section{Background}
Electroweak theory couples the baryon (B) and the lepton (L) numbers to the Chern-Simons number of the weak gauge field through the axial anomaly.  At temperatures higher than the electroweak phase transition, the rate of Chern-Simons number fluctuations -- the sphaleron rate -- has a nonzero value, whereas at lower temperatures it is exponentially suppressed and, when the Higgs field expectation value $v \gg T$, the rate is negligible.  In electroweak baryogenesis scenarios \cite{Kuzmin:1985mm} the baryon number of the Universe is generated during the electroweak phase transition.  However, this scenario does not work in the Standard Model: it requires a strongly first order phase transition,
whereas the Standard Model has a smooth crossover \cite{Kajantie:1996mn}. Further, the CP violation
in the Standard Model is not sufficient to drive baryon number generation.

Nevertheless, the sphaleron rate during the electroweak crossover in the Standard Model is 
relevant for some Leptogenesis scenarios: in these scenarios lepton asymmetry is converted into baryon
asymmetry through sphaleron transitions.  If the lepton asymmetry is generated just before or during 
the electroweak phase transition, how the sphaleron rate shuts off has an effect on the generated baryon number.
The sphaleron rate has been studied in the broken phase before, but either with unphysical Higgs masses \cite{Moore:1998swa,Moore:2000jw,Tang:1996qx} or not very deeply in the broken phase \cite{Moore:1998swa}.

In the electroweak theory, the gauge field vacua are labeled by the Chern-Simons number
\begin{equation}
n_{CS} = \int d^3 x \ j_{CS}^0 = -\frac{g^2}{64 \pi}\int d^3x \ \epsilon^{ijk} \textrm{Tr}\left(A_i F_{jk} + i \frac{g}{3} A_i A_j A_k \right).
\end{equation}
The Chern-Simons current $j_{CS}^{\mu}$ is in turn related through the axial anomaly to the baryon- and lepton-number currents
\begin{equation}
\partial_{\mu} (j_B^{\mu}+j_L^{\mu}) = n_g \left(\frac{g^2}{16 \pi^2}\epsilon^{\alpha \beta \mu \nu} A^a_{\alpha \beta} A^a_{\mu \nu} \right),
\end{equation}
by 
\begin{equation}\label{baryoncurrent}
\partial_{\mu} \ j_B^{\mu} = n_g \ \partial_{\mu} \ j_{CS}^{\mu},
\end{equation}
where the U(1) part of the theory is omitted.
Transitions between vacua are possible by surmounting the potential barrier through sphaleron transitions. The sphaleron rate is strongly suppressed at low temperatures, where the potential barrier is high. At temperatures above the EWPT, though, transitions among vacua are made possible through thermal fluctuations because there is no longer any potential barrier.
Each transition changes $n_{CS}$ by one unit and therefore violates the baryon number by $n_g$ $=$ 3
\begin{displaymath}
B(t_f)-B(t_i) = n_g \ [n_{CS}(t_f)-n_{CS}(t_i)]
\end{displaymath}
thus providing a source of Baryogenesis.

In previous works, the sphaleron rate has been studied at the energy range of the electroweak phase transition either in the symmetric phase with lattice simulations \cite{Ambjorn:1990pu} and semiclassical methods \cite{Philipsen:1995sg}, or in the broken phase with both perturbative calculations \cite{Burnier:2005hp} and on the lattice \cite{Moore:1998swa,Krasnitz:1993mt}.

In this work we unify these two pictures and find the overall behavior of the sphaleron rate from the symmetric phase to the broken one, passing through the electroweak crossover. Our results are compared to analytical estimates both in the broken and symmetric phases \cite{Burnier:2005hp}. 
\section{Theory on the lattice}
The thermodynamics of the 4-dimensional electroweak theory is studied in 3 dimensions by means of dimensional reduction \cite{Kajantie:1995dw}, a perturbative technique that gives the correspondence between 4D and 3D parameters. The result is a SU(2) effective theory with the Higgs field $\phi$ and gauge field $A_{\mu}$ ($F_{ij}$)
\begin{equation}
L = \frac{1}{4} F^a_{ij} F^a_{ij} + (D_i\phi)^{\dagger}(D_i\phi)+m_3^2 \phi^\dagger \phi + \lambda_3 (\phi^\dagger \phi)^2,
\end{equation}
and 3D effective parameters $g_3^2$, $\lambda_3$ and $m_3^2$. 

B\"odeker showed \cite{Bodeker:1998hm} that at leading order in log(1/$g$) the time evolution of this effective SU(2) Higgs model is governed by Langevin dynamics. The latter, though, is very slow on the lattice and can be substituted by any other dissipative procedure, e.~g.~heat bath. 
One heat-bath sweep through the lattice corresponds to the real-time step \cite{Moore:2000jw}
\begin{equation}
\Delta t = \frac{a^2 \ \sigma_{el}}{4}, 
\end{equation}
where
\begin{equation}\label{Moore3.9}
\sigma^{-1}_{\textrm{el}} = \frac{3}{m^2_D}\gamma , \quad \textrm{with} \quad \gamma= \frac{N g^2 T}{4 \pi} \left[ln\frac{m_D}{\gamma}+3.041 \right]
\end{equation}
is the non-abelian color conductivity, which quantifies the current response to infrared external fields, $N$ is the dimension of the SU(N) gauge group, and $m_D$ is the Debye mass, determining the length scale $l_D$ $\sim$ $1/m_D$ $\sim$ $1/gT$. We made use of a 32$^3$ lattice, with $\beta_G \equiv \frac{4}{g_3^2 a} =$ 9, where $g_3$ is the 3D gauge coupling and $a$ the lattice spacing. In real-time simulations, for each mass and temperature pair, we computed 4 trajectories for every 1000 initial configurations.
\section{Methods}
\vspace*{-3 mm}
In the symmetric phase we make use of canonical MC simulations and approach the broken phase. 
At very low temperatures, the rate is highly suppressed and canonical methods do not work anymore. 
We need multicanonical methods, which calculate a weight function that compensates the high potential barrier between the vacua, thus allowing transitions.
The exact value of the sphaleron rate 
\begin{equation}
\Gamma \equiv \lim_{t\rightarrow \infty} \frac{\langle(n_{CS}(t)-n_{CS}(0))^2\rangle}{V \ t}
\end{equation}
is obtained, in the broken phase, through a method similar to the one used in \cite{Moore:1998swa,Moore:2000jw}. 

\begin{description}
\item[]\textbf{a.} First we fix the order parameter ($n^*_{CS}$ $=$ $1/2$ in our case) which separates one vacuum from the neighbouring one. 
\item[]\textbf{b.} We calculate the probability for $n_{CS}$ to be in the small interval $n^*_{CS} \pm \epsilon/2$. This can be achieved only with multicanonical methods, as the probability $P_{\epsilon}$ of being on top of the barrier is extremely small.
\item[]\textbf{c.} Then the probability $P_{\epsilon}$ is transformed into a flux by multiplying it with $\langle dn_{CS} / dt \rangle / \epsilon$. This is calculated by taking initial configurations in the interval $\epsilon$, performing real-time simulations 
and keeping track of the $n_{CS}$ value after some time $dt$.
\item[]\textbf{d.} Finally, we calculate the \textsl{dynamical prefactor}
\begin{equation}
\textbf{d} = \sum_{sample} \frac{\delta}{\# \ \textrm{crossings}}
\end{equation}
which is a measure of the fraction of the crossings which lead to a permanent change in $n_{CS}$. $\delta$ is 0 for configurations that return to the initial vacuum and 1 if the initial and final vacuum are different. The initial configurations are chosen to be in $n^*_{CS}$ $\pm$ $\epsilon / 2$ 
and the real-time evolution is performed forward and backwards in time.
\item[]\textbf{e.} The sphaleron rate is then
\begin{equation}
\Gamma \equiv \frac{P(\mid n_{CS}-n^*_{CS} \mid < \epsilon/2) } {\epsilon \ P(\mid n_{CS} < n^*_{CS}\mid)} \left\langle  \mid \frac{dn_{CS}}{dt} \mid \right\rangle  \times \textbf{d}.
\end{equation}
\end{description}

\section{Results}
\begin{figure}
\vspace*{-8 mm}
\centerline{
\includegraphics[width=.43\linewidth]{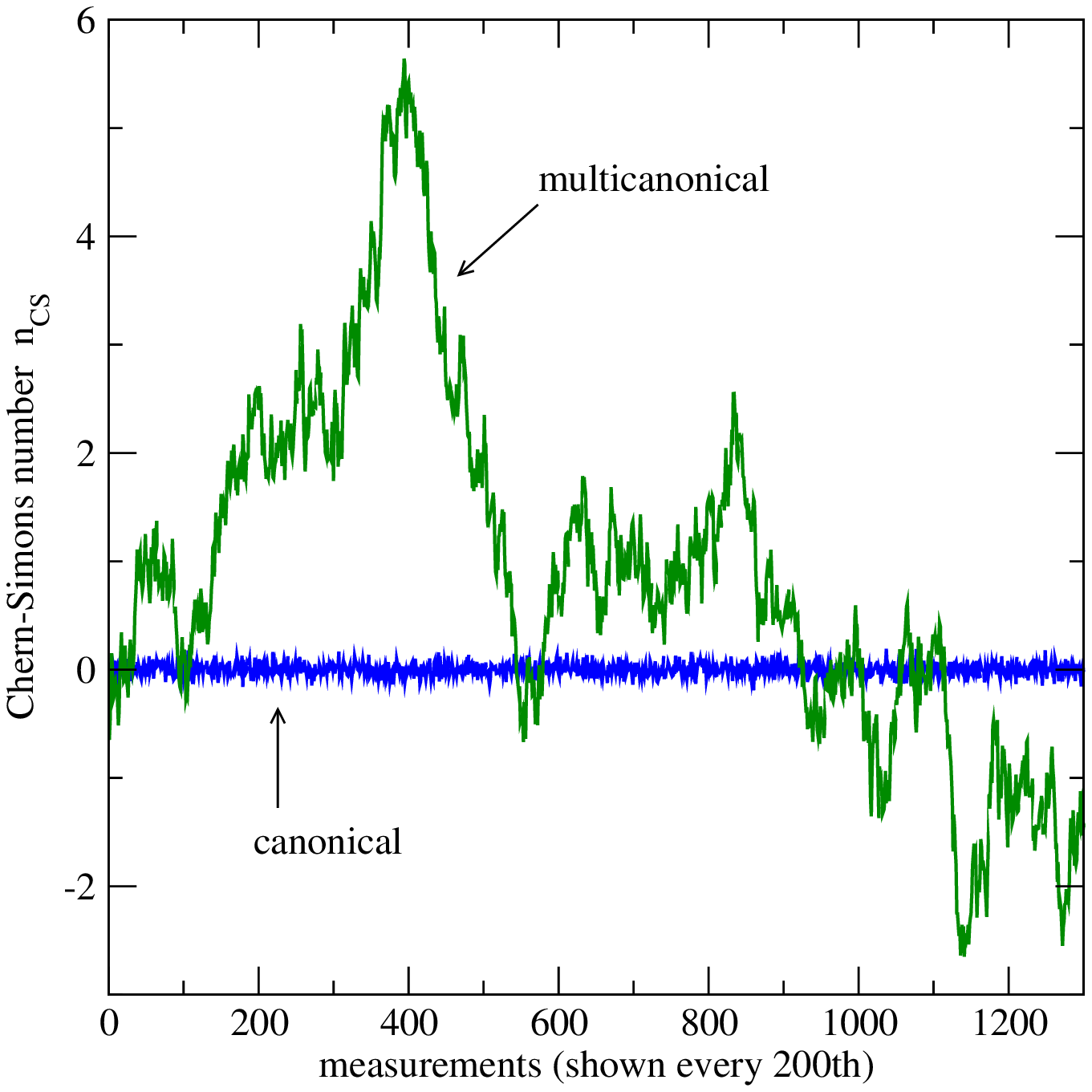} 
\includegraphics[width=.43\linewidth]{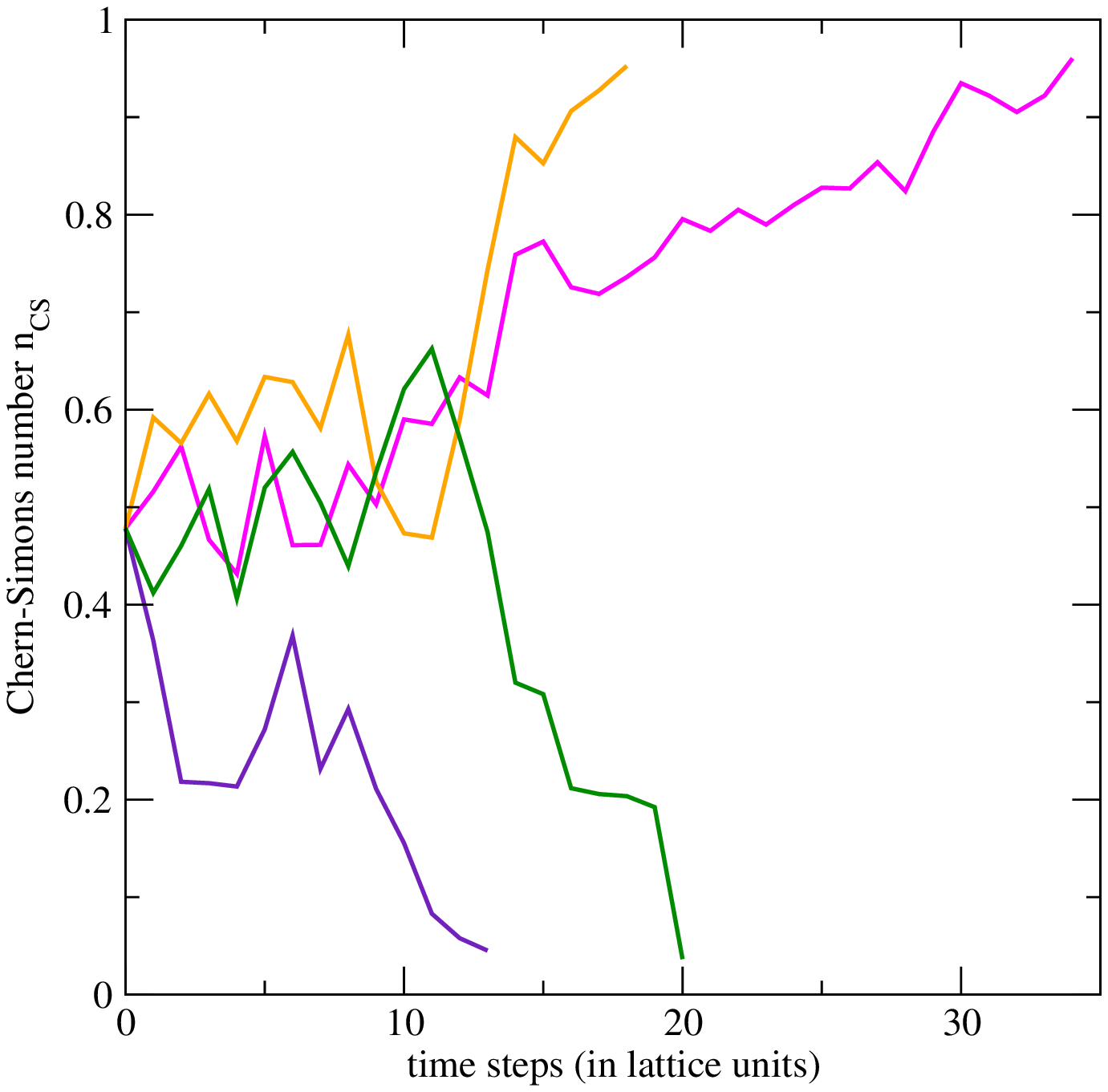} \\
}
\caption[a]{\small{Left: The Chern-Simons number evolution below the critical temperature ($m_H =$ 115 GeV, $T =$~142 GeV) in canonical and multicanonical simulations. The transition rate in this plot is not related to the real time rate, but shows the efficiency of the probability distribution measurement. Right: A set of heat-bath trajectories originating from the same configuration. Gluing together any two of these produces a trajectory, which corresponds to a sphaleron transition if the two end-points are in different minima.} 
}
\label{csfig}
\end{figure}
\begin{figure}
\vspace*{-4 mm}
\centerline{
\epsfig{file=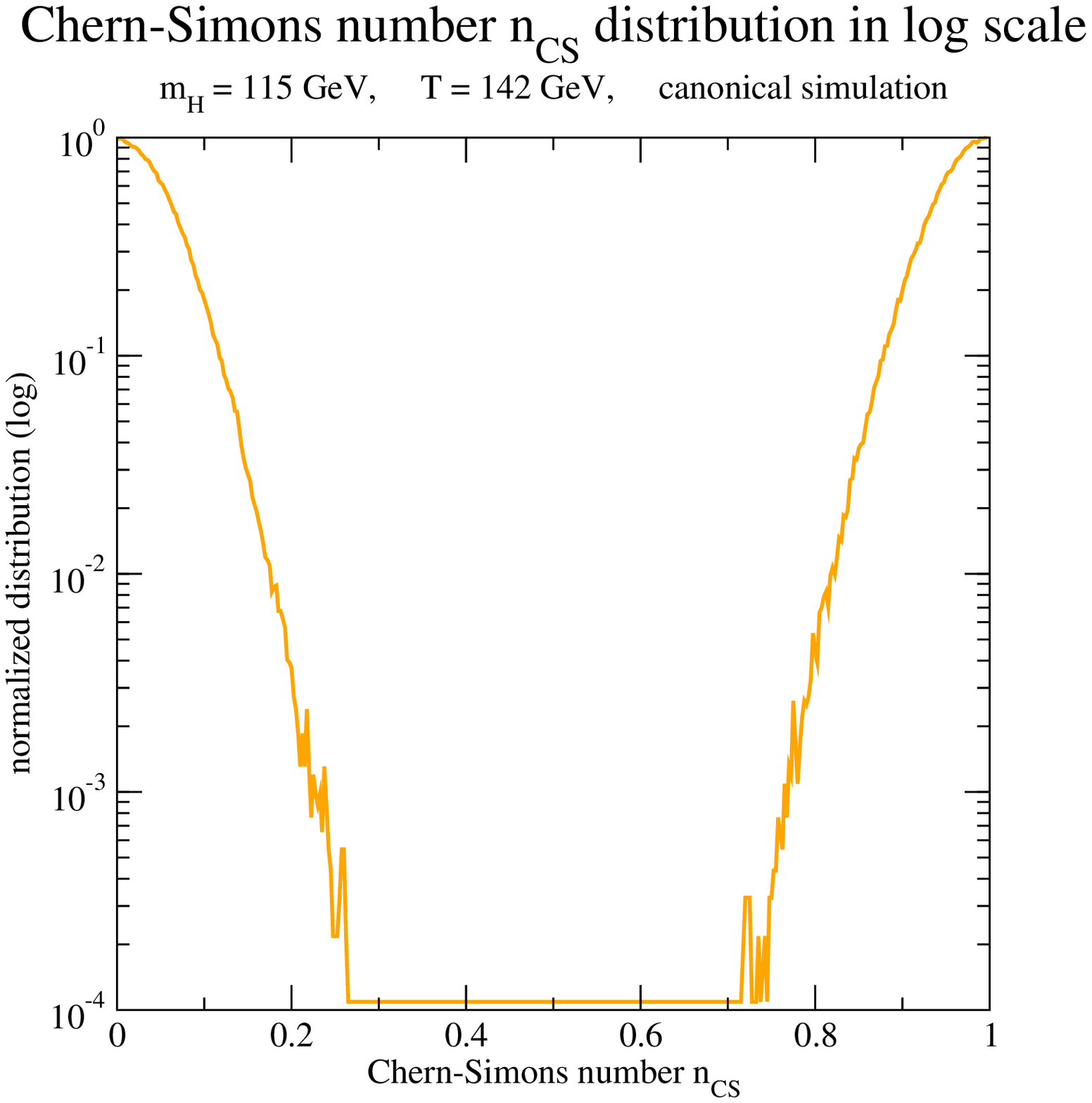,width=0.42\linewidth,clip=} 
\epsfig{file=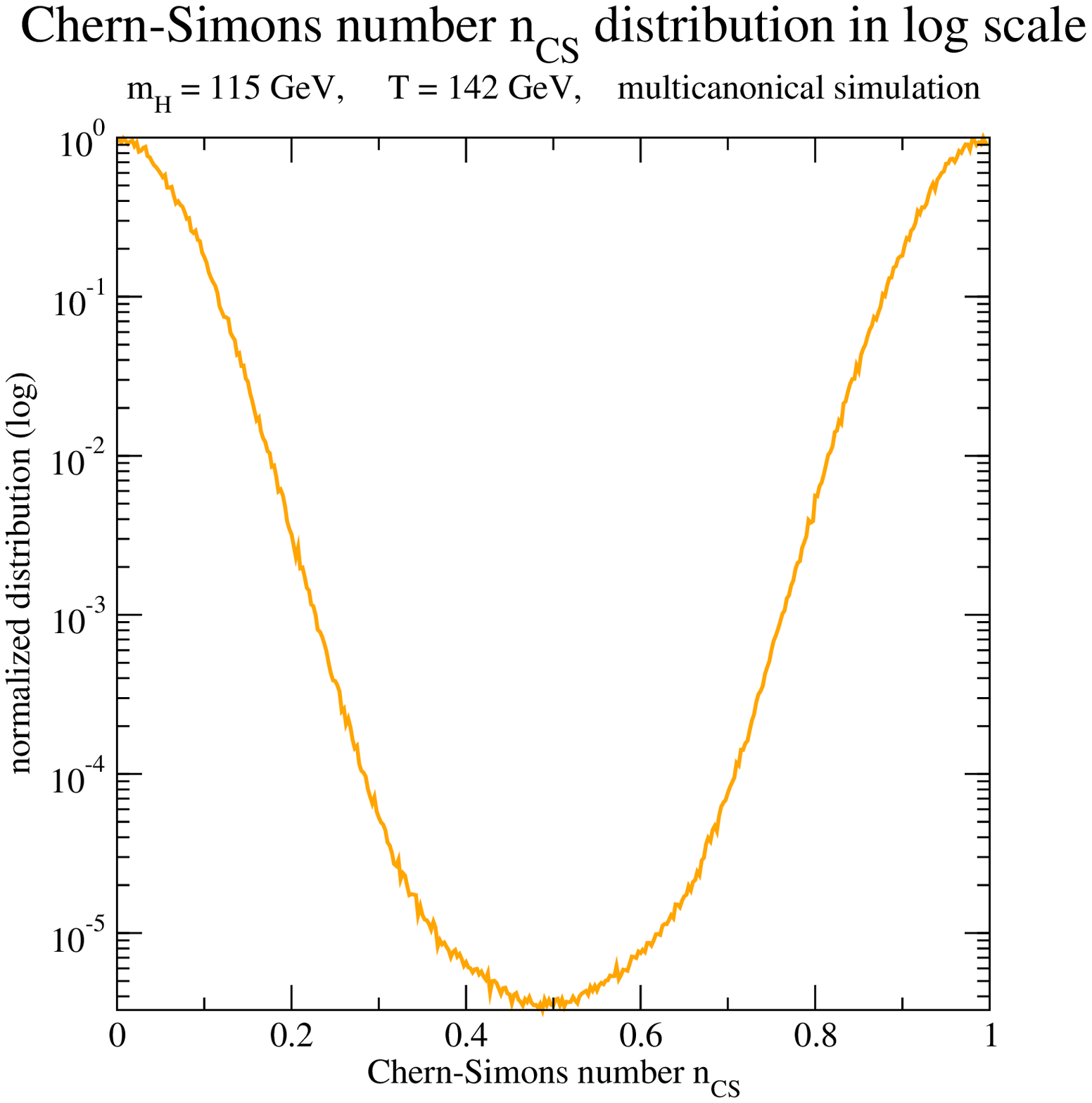,width=0.42\linewidth,clip=} \\
}
\caption[a]{\small{The probability distributions of Chern-Simons number in the deep broken phase in canonical (left) and multicanonical simulations (right).}}
\label{plog_both}
\end{figure}

Figure \ref{csfig} (left)
shows the efficiency of the multicanonical method at low temperatures. For the Higgs mass of 115 GeV and the temperature of 142 GeV, we see that in the canonical simulation, no transitions happen, while in the multicanonical run we have a random walk in the adjusted potential, where we have compensated for the statistical suppression by the weight function $W$.
This can also be seen in the probability distributions these simulations produce (Figure \ref{plog_both}) for the same $T$ $=$ 142 GeV.

The multicanonical weight function $W$ thus permits sampling with constant probability, being the conversion factor between multicanonical and physical probability
\begin{equation}
P_{muca} \varpropto exp[W] \ P_{can}.
\end{equation}

Figure \ref{csfig} (right) shows several real-time heat-bath trajectories from the same initial configuration.
Each trajectory crosses a different number of times the least-probable interval $\epsilon$ on the top of the barrier, and ends either back into the initial vacuum or into the adjacent one. 

\begin{figure}[!h]
\vspace*{-7.5 mm}
\begin{center}
\includegraphics[width=.495\linewidth]{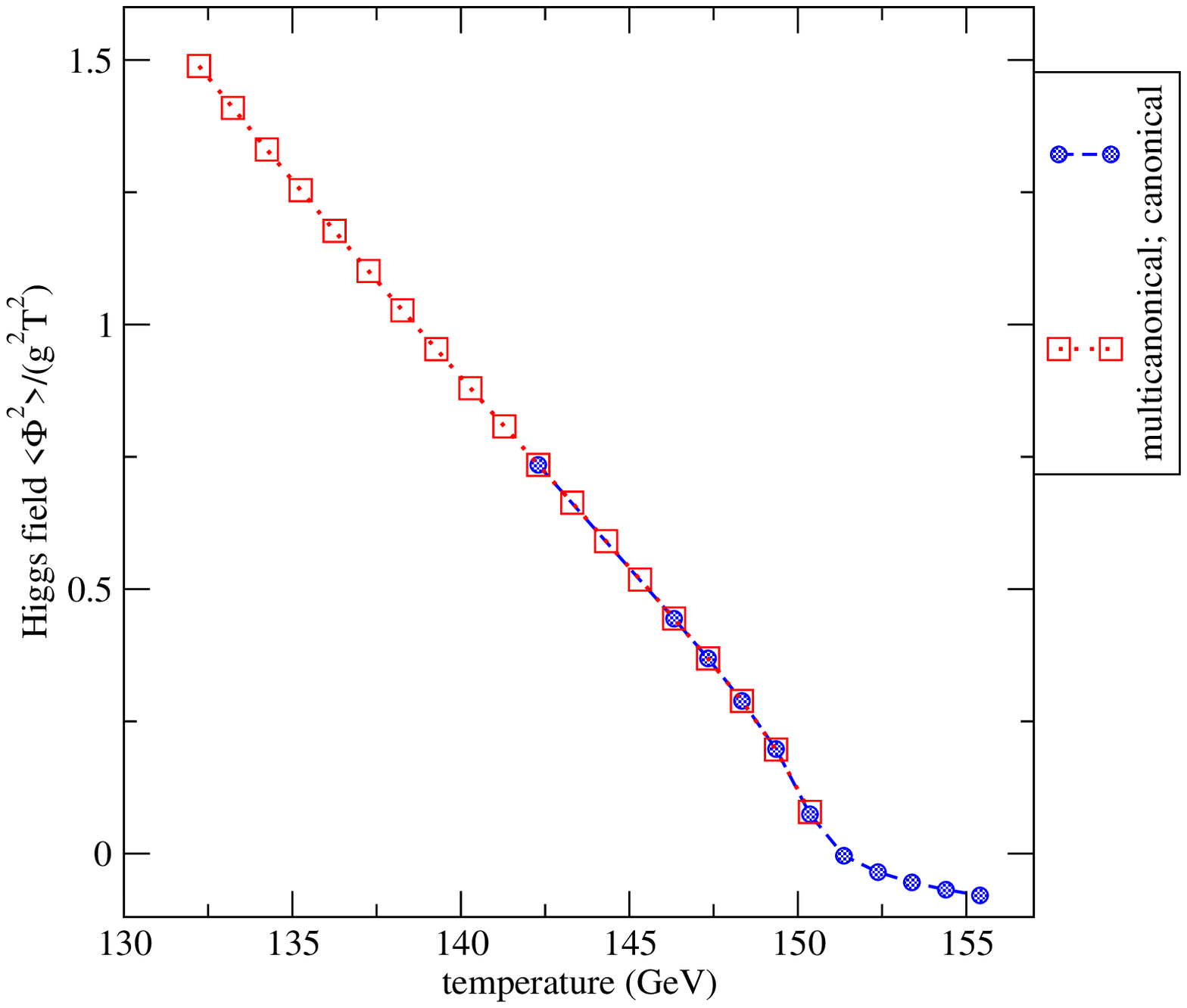}
\includegraphics[width=.495\linewidth]{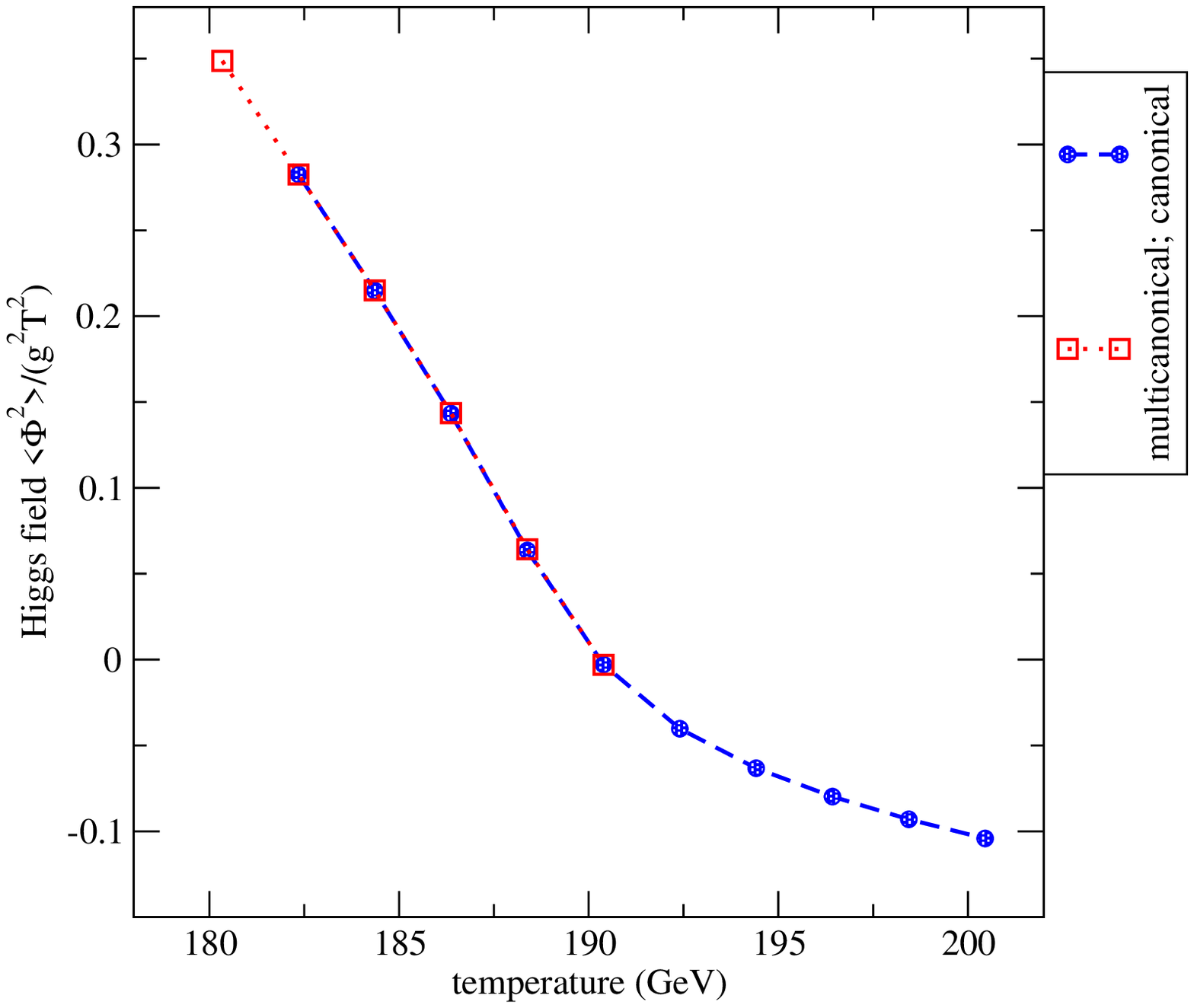}
\caption{\small{The Higgs expectation value $\langle\phi^2\rangle$ for Higgs masses of 115 GeV (left) and 160 GeV (right) as a function of temperature. The high-temperature canonical and low-temperature multicanonical results match beautifully in the transition region.}}
\label{phisq}
\end{center}
\end{figure}

In  Figure \ref{phisq} we show the Higgs field expectation value $\langle\phi^2\rangle$ for both masses (115 GeV, left, and 160 GeV, right) as a function of temperature. We notice a perfect match between the canonical and multicanonical results and a smooth transition from the symmetric to the broken phase.

\begin{figure}[!h]
\vspace*{-5 mm}
\begin{center}
\includegraphics[width=.58\linewidth]{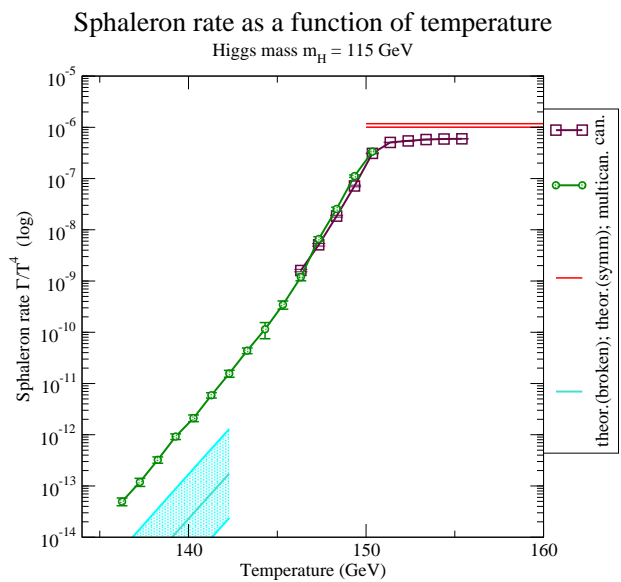}\\
\includegraphics[width=.58\linewidth]{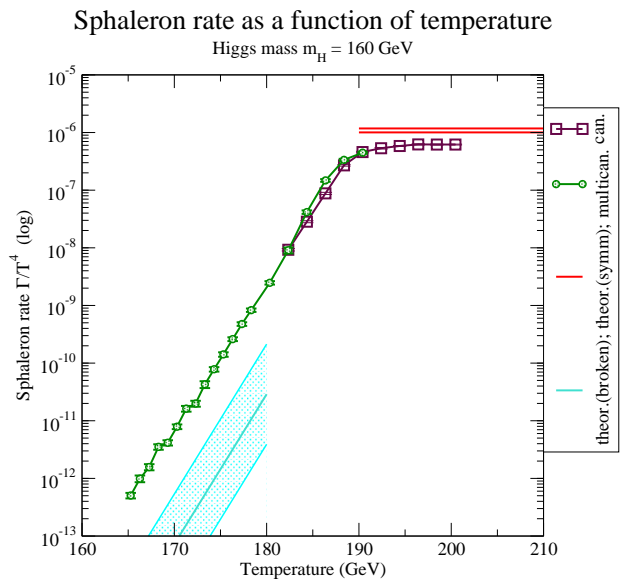}
\caption{\small{The sphaleron rate for a Higgs mass of 115 GeV (above) and 160 GeV (below). The high-temperature canonical and low-temperature multicanonical results again match very well in the transition region. Also shown are previous high-temperature estimates (top, horizontal line) and perturbative calculations in the low-temperature phase (bottom, wide band) from \cite{Burnier:2005hp}.}}
\label{sphrate}
\end{center}
\end{figure}

The sphaleron rate $\Gamma / T^4$ is shown in Figure \ref{sphrate} for $m_H =$ 115 GeV and 160 GeV, with the theoretical curves obtained separately for the broken and symmetric phases, through perturbative calculations in \cite{Burnier:2005hp}. 

\section{Conclusion}
We improved the previous estimates for the sphaleron rate and determined its behaviour from the symmetric to the broken phase, through the electroweak crossover. Our results are in agreement with previous estimates in the symmetric phase. In the broken phase we notice that the slope of our curve is the same as in the analytic one \cite{Burnier:2005hp}. We however note a discrepancy of up to two orders of magnitude in the size of the rate, although part of the shift in temperature may be explained in terms of renormalization constants. 

Even though the Standard Model has a too weak source of CP-violation in the quark sector, Baryogenesis might still be viable through lepton number violating processes. The sphaleron rate plays an important role in Leptogenesis, as the conversion of lepton to baryon number depends on it, and it is therefore important to know its size rather accurately. 

\acknowledgments
This work is supported by the Academy of Finland grants 1134018 and 114371.
M.D. also acknowledges support from the Magnus Ehrnrooth foundation.
The computations have been performed at the Finnish IT center for Science (CSC).

\end{document}